\documentclass[]{article}

\usepackage{siunitx}
\usepackage{graphicx,rotating,multirow}
\usepackage{bm,multirow}
\usepackage{amsmath,amssymb,color,mathrsfs,citesort}

\newcommand{\AJP}{ {\em Am. J. Phys.} }

\newcommand{\AnM}{ {\em Annals Math. }}

\newcommand{\APB}{ {\em Ann. Phys. (Berlin)} }
\newcommand{\APNY}{ {\em Ann. Phys. (N.Y.)} }

\newcommand{\CMP}{ {\em Commun. Math. Phys.} }

\newcommand{\CRA}{ {\em C. R. Acad. Sci. Ser. A} }
\newcommand{\EJP}{ {\em Eur. J. Phys.} }
\newcommand{\EPJD}{ {\em Eur. Phys. J. D} }
\newcommand{\EPL}{ {\em Europhys. Lett.} }
\newcommand{\EPJST}{ {\em Eur. Phys. J. Spec. Top.} }

\newcommand{\IJQC}{ {\em Int. J. Quantum Chem.} }

\newcommand{\JMC}{ {\em J. Math. Chem.} }
\newcommand{\JMP}{ {\em J. Math. Phys.} }
\newcommand{\jpa}{ {\em J. Phys. A} }

\newcommand{\JSM}{ {\em J. Stat. Mech.} }
\newcommand{\JSP}{ {\em J. Stat. Phys.} }

\newcommand{\NJP}{ {\em New J. Phys.} }
\newcommand{\NL}{ {\em Nature (London)} }

\newcommand{\PA}{ {\em Physica A} }

\newcommand{\PLA}{ {\em Phys. Lett. A} }
\newcommand{\PNAS}{ {\em P. Natl. Acad. Sci. USA }}

\newcommand{\PRA}{ {\em Phys. Rev. A} }

\newcommand{\PRe}{ {\em Phys. Rep.} }

\newcommand{\RMP}{ {\em Rev. Mod. Phys.} }
\newcommand{\RMS}{ {\em Russ. Math. Surv.} }

\newcommand{\UMN}{ {\em Usp. Mat. Nauk} }

\begin{document}
\title{R\'{e}nyi and Tsallis entropies of the Dirichlet and Neumann one-dimensional quantum wells}
\author{O. Olendski\footnote{{Department of Applied Physics and Astronomy, University of Sharjah, P.O. Box 27272, Sharjah, United Arab Emirates}}}
\maketitle
\begin{abstract}
A comparative analysis of the Dirichlet and Neumann boundary conditions (BCs) of the one-dimensional (1D) quantum well extracts similarities and differences of the R\'{e}nyi $R(\alpha)$ as well as  Tsallis $T(\alpha)$ entropies between these two geometries. It is shown, in particular, that for either BC the dependencies of the R\'{e}nyi position components on the parameter $\alpha$ are the same for all orbitals but the lowest Neumann one for which the corresponding functional $R$ is not influenced by the variation of $\alpha$. Lower limit $\alpha_{TH}$ of the semi infinite range of the dimensionless R\'{e}nyi/Tsallis coefficient where {\em momentum} entropies exist crucially depends on the {\em position} BC and is equal to one quarter for the Dirichlet requirement and one half for the Neumann one. At $\alpha$ approaching this critical value, the corresponding momentum functionals do diverge. The gap between the thresholds $\alpha_{TH}$ of the two BCs causes different behavior of the R\'{e}nyi uncertainty relations as functions of $\alpha$. For both configurations, the lowest-energy level at $\alpha=1/2$ does saturate either type of the entropic inequality thus confirming an earlier surmise about it. It is also conjectured that the threshold $\alpha_{TH}$ of one half is characteristic of any 1D non-Dirichlet system. Other properties are discussed and analyzed from the mathematical and physical points of view.
\end{abstract}

\section{Introduction}
Rapidly growing interest from physicists, mathematicians, chemists and other researchers to the study of the quantum information measures is stimulated by the fact that these functionals  of the one-particle position $\rho({\bf r})$ and wave vector $\gamma({\bf k})$ densities describe various important properties of miscellaneous nano-sized confined structures; for example, Shannon entropies \cite{Shannon1}
\begin{subequations}\label{Shannon1}
\begin{align}\label{Shannon1_X}
S_{\rho_n}&=-\int_{\mathcal{D}_\rho^l}\rho_n({\bf r})\ln\rho_n({\bf r})d{\bf r}\\
\label{Shannon1_P}
S_{\gamma_n}&=-\int_{\mathcal{D}_\gamma^l}\gamma_n({\bf k})\ln\gamma_n({\bf k})d{\bf k}
\end{align}
\end{subequations} 
characterize localization/delocalization of the $n$-th bound orbital in the corresponding $l$-dimensional position (subscript $\rho$) or momentum (subscript $\gamma$) space; Fisher informations \cite{Fisher1,Frieden1}
\begin{subequations}\label{GeneralFisher1}
\begin{align}\label{GeneralFisherX1}
I_{\rho_n}&=\int_{\mathcal{D}_\rho^l}\rho_n({\bf r})\left|{\bm\nabla}\ln\rho_n({\bf r})\right|^2d{\bf r}=\int_{\mathcal{D}_\rho^l}\frac{\left|{\bm\nabla}\rho_n({\bf r})\right|^2}{\rho_n({\bf r})}d{\bf r}\\
\label{GeneralFisherK1}
I_{\gamma_n}&=\int_{\mathcal{D}_\gamma^l}\gamma_n({\bf k})\left|{\bm\nabla}\ln\gamma_n({\bf k})\right|^2d{\bf k}=\int_{\mathcal{D}_\gamma^l}\frac{\left|{\bm\nabla}\gamma_n({\bf k})\right|^2}{\gamma_n({\bf k})}\,d{\bf k}
\end{align}
\end{subequations}
due to the presence of the gradients provide a quantitative estimation of the oscillating structure of each probability distribution, and Onicescu energies \cite{Onicescu1}
\begin{subequations}\label{Onicescu1}
\begin{align}\label{Onicescu1_X}
O_{\rho_n}=\int_{\mathcal{D}_\rho^l}\rho_n^2({\bf r})d{\bf r}\\
\label{Onicescu1_K}
O_{\gamma_n}=\int_{\mathcal{D}_\gamma^l}\gamma_n^2({\bf k})d{\bf k}
\end{align}
\end{subequations}
yield the numbers that categorize deviations of the densities from the uniform ones. Here, positive integer $n$ counts all bound orbitals in the ascending order of their energies and densities are squared magnitudes of the corresponding waveforms $\Psi_n({\bf r})$ and $\Phi_n({\bf k})$:
\begin{subequations}\label{Densities1}
\begin{align}
\label{DensityX1}
\rho_n({\bf r})&=\left|\Psi_n({\bf r})\right|^2\\
\label{DensityP1}
\gamma_n({\bf k})&=\left|\Phi_n({\bf k})\right|^2,
\end{align}
\end{subequations}
which are orthonormalized:
\begin{equation}\label{OrthoNormality1}
\int_{\mathcal{D}_\rho^l}\Psi_{n'}^\ast({\bf r})\Psi_n({\bf r})d{\bf r}=\int_{\mathcal{D}_\gamma^l}\Phi_{n'}^\ast({\bf k})\Phi_n({\bf k})d{\bf k}=\delta_{nn'},
\end{equation}
with $\delta_{nn'}=\left\{\begin{array}{cc}
1,&n=n'\\
0,&n\neq n'
\end{array}\right.$ being a Kronecker delta, $n,n'=1,2,\ldots$, and related to each other via the Fourier transform:
\begin{subequations}\label{Fourier1}
\begin{align}\label{Fourier1_1}
\Phi_n({\bf k})&=\frac{1}{(2\pi)^{l/2}}\int_{\mathcal{D}_\rho^l}\Psi_n({\bf r})e^{-i{\bf kr}}d{\bf r},\\
\label{Fourier1_2}
\Psi_n({\bf r})&=\frac{1}{(2\pi)^{l/2}}\int_{\mathcal{D}_\gamma^l}\Phi_n({\bf k})e^{i{\bf rk}}d{\bf k}.
\end{align}
\end{subequations}
In all these equations, integrations are carried out over the whole region $\mathcal{D}_\rho^l$ or $\mathcal{D}_\gamma^l$ where the function $\Psi({\bf r})$ or $\Phi({\bf k})$ is defined; for example, for the particle of mass $m^*$ in the one-dimensional (1D), $l=1$, Dirichlet ($D$) or Neumann ($N$) quantum wells studied below, the position domain $\mathcal{D}_\rho^1$ is strictly the finite interval $|x|<a/2$, with $a$ being the width of the structure, 
\begin{equation}\label{Domain1}
\mathcal{D}_\rho^1\left(-\frac{\hbar^2}{2m^*}\frac{d^2}{dx^2}\right)=\left\{\Psi,\Psi'\in\mathcal{L}^2(-a/2,+a/2),\left\{\begin{array}{c}
\Psi^D(\pm a/2)=0\\
{\Psi^N}'(\pm a/2)=0
\end{array}\right\}\right\},
\end{equation}
whereas the wave vector $k$ spans the whole real axis, $\mathcal{D}_\gamma^1=\mathbb{R}^1$.

Contrary to the above functionals, R\'{e}nyi \cite{Renyi1,Renyi2}
\begin{subequations}\label{FunctionalsR1}
\begin{align}\label{RenyiX1}
R_{\rho_n}(\alpha)&=\frac{1}{1-\alpha}\ln\!\left(\int_{\mathcal{D}_\rho^l}\rho_n^\alpha({\bf r})d{\bf r}\right)\\
\label{RenyiP1}
R_{\gamma_n}(\alpha)&=\frac{1}{1-\alpha}\ln\!\left(\!\int_{\mathcal{D}_\gamma^l}\gamma_n^\alpha({\bf k})d{\bf k}\right)
\end{align}
\end{subequations}
and Tsallis \cite{Tsallis1}
\begin{subequations}\label{FunctionalsT1}
\begin{align}\label{TsallisX1}
T_{\rho_n}(\alpha)&=\frac{1}{\alpha-1}\left(1-\int_{\mathcal{D}_\rho^l}\rho_n^\alpha({\bf r})d{\bf r}\right)\\
\label{TsallisP1}
T_{\gamma_n}(\alpha)&=\frac{1}{\alpha-1}\left(1-\int_{\mathcal{D}_\gamma^l}\gamma_n^\alpha({\bf k})d{\bf k}\right),
\end{align}
\end{subequations}
entropies are determined not only by the charge distributions but depend also on the non-negative dimensionless coefficient $\alpha$, which can be construed as a factor describing the reaction of the system to its deviation from the equilibrium. At $\alpha=1$, the l'H\^{o}pital's rule transforms both functionals into the Shannon entropies, Equations~\eqref{Shannon1}. Onicescu energies can be expressed through them too:
\begin{subequations}\label{Onicescu2}
\begin{align}
\label{Onicescu2_X}
O_{\rho_n}&=e^{-R_{\rho_n}(2)}=1-T_{\rho_n}(2)\\
\label{Onicescu2_P}
O_{\gamma_n}&=e^{-R_{\gamma_n}(2)}=1-T_{\gamma_n}(2).
\end{align}
\end{subequations}
General physical and mathematical properties and meaning of these one-parameter functionals are thoroughly described in many sources; so, here let us just mention that Equations~\eqref{Shannon1}--\eqref{Onicescu1} and \eqref{FunctionalsR1}, \eqref{FunctionalsT1} for the continuous random variables were obtained as a generalization of their discrete counterparts, which are dimensionless and strictly positive. However, Shannon and R\'{e}nyi entropies defined above are measured in units of the logarithms of length and can take negative values whereas Tsallis components represent a sum of the dimensionless unity and the quantity measured in units of the distance raised to positive or negative power of $l(1-\alpha)$.

Position and momentum components of the R\'{e}nyi as well as Tsallis entropy for the same orbital are not independent from each other; namely, Sobolev inequality of the Fourier transform \cite{Beckner1}
\begin{equation}\label{Sobolev1}
\left(\frac{\alpha}{\pi}\right)^{l/(4\alpha)}\left[\int_{\mathcal{D}_\rho^l}\rho_n^\alpha({\bf r})d{\bf r}\right]^{1/(2\alpha)}\geq\left(\frac{\beta}{\pi}\right)^{l/(4\beta)}\left[\int_{\mathcal{D}_\gamma^l}\gamma_n^\beta({\bf k})d{\bf k}\right]^{1/(2\beta)},
\end{equation}
where parameters $\alpha$ and $\beta$ are conjugated as
\begin{equation}\label{RenyiUncertainty1}
\frac{1}{\alpha}+\frac{1}{\beta}=2,
\end{equation}
leads directly to the interrelation between the Tsallis entropies \cite{Rajagopal1}:
\begin{equation}\label{TsallisInequality1}
\left(\frac{\alpha}{\pi}\right)^{l/(4\alpha)}\!\!\left[1+(1-\alpha)T_{\rho_n}(\alpha)\right]^{1/(2\alpha)}\geq\left(\frac{\beta}{\pi}\right)^{l/(4\beta)}\!\!\left[1+(1-\beta)T_{\gamma_n}(\beta)\right]^{1/(2\beta)}, \end{equation}
which is obviously saturated at $\alpha=\beta=1$ around which point these inequalities degenerate to \cite{Olendski3}
\begin{equation}\label{TsallisInequality2}
\frac{1+\left[-2S_{\rho_n}+l(1+\ln\pi)\right](\alpha-1)/4}{\pi^{l/4}}\geq\frac{1+\left[2S_{\gamma_n}-l(1+\ln\pi)\right](\alpha-1)/4}{\pi^{l/4}},\,\alpha\rightarrow1.
\end{equation}
Invoking Shannon uncertainty relation, Equation~\eqref{ShannonInequality} below, one sees that, in addition to the requirement from Equation~\eqref{RenyiUncertainty1}, an additional constraint
\begin{equation}\label{Sobolev2}
\frac{1}{2}\leq\alpha\leq1
\end{equation}
is imposed onto the Sobolev, Equation~\eqref{Sobolev1}, and Tsallis, Equation~\eqref{TsallisInequality1}, inequalities. On the examples of several 1D systems, it was shown recently \cite{Olendski1} that Equations~\eqref{Sobolev1} and \eqref{TsallisInequality1} turn into equality also for the ground state at $\alpha=1/2$ and it was conjectured that this property holds true for any $l$-dimensional system when either side of these inequalities turns to $\Phi_1({\bf 0})$. Taking the logarithm of both sides of Equation~\eqref{Sobolev1}, one derives the R\'{e}nyi uncertainty relation \cite{Bialynicki1,Zozor1}
\begin{equation}\label{RenyiUncertainty2}
R_{\rho_n}(\alpha)+R_{\gamma_n}(\beta)\geq-\frac{l}{2}\left(\frac{1}{1-\alpha}\ln\frac{\alpha}{\pi}+\frac{1}{1-\beta}\ln\frac{\beta}{\pi}\right),
\end{equation}
for which the constraint from Equation~\eqref{Sobolev2} has been waived. At $\alpha=1/2$, when the right-hand side of Equation~\eqref{RenyiUncertainty2} turns to $l\ln(2\pi)$, the ground level converts this relation into the equality too \cite{Olendski1} whereas for $\alpha=\beta=1$ for any state it degenerates into its Shannon counterpart \cite{Bialynicki2,Beckner2}:
\begin{equation}\label{ShannonInequality}
S_{\rho_n}+S_{\gamma_n}\geq l(1+\ln\pi).
\end{equation}
Observe that Equations~\eqref{RenyiUncertainty2} and \eqref{ShannonInequality} contain dimensionless scale independent scalars since the logarithms of length enter into the expressions for the position $R_{\rho_n}$ and momentum $R_{\gamma_n}$ components with the opposite signs. Let us note also that in the presence of the magnetic fields the orbital that at $\alpha=1/2$ saturates the R\'{e}nyi, Equation~\eqref{RenyiUncertainty2}, and Tsallis, Equation~\eqref{TsallisInequality1}, inequalities is not necessarily the lowest energy level, as was shown for the 2D ring \cite{Olendski3}. A study of the entropic uncertainty relations \cite{Wehner1,Jizba1,Coles1,Hertz1,Wang1} is an essential endeavor both from a fundamental point of view with respect to their role in quantum foundations as well as their miscellaneous applications, first of all, in information theory.

Both one-parameter functionals $R(\alpha)$ and $T(\alpha)$ are widely employed in various branches of human activity. Recent analysis \cite{Olendski1} provided a by no means complete mini review of the very diverse fields where these two entropies are successfully implemented. Relevant to the nanotechnology research, one has to mention cutting-edge experiments that succeeded in direct estimation of the R\'{e}nyi entanglement entropies with $\alpha=2$ of the Bose-Einstein condensates of the interacting atoms and ions \cite{Islam1,Kaufman1,Brydges1}. From point of view of quantum chemistry, it is essential to understand the properties of the R\'{e}nyi and Tsallis measures for three fundamental structures:  infinite potential well \cite{Belloni1}, harmonic oscillator and hydrogen atom. In the last four years or so, corresponding research  addressed many features of these entropies for the latter two systems \cite{Aptekarev3,Aptekarev2,Toranzo1,Toranzo2,Dehesa1,PuertasCenteno4,Belega1,PuertasCenteno3,PuertasCenteno1,PuertasCenteno2,Mukherjee1}, including their multidimensional generalizations. Analysis of the Dirichlet quantum well (i.e., a flat 1D structure of finite spatial extent at the borders of which the wave function turns to zero) lacks such completeness. Whereas an exact expression for the position R\'{e}nyi entropy was derived quite long ago \cite{SanchezRuiz1}, momentum components were found for the integer coefficient $\alpha$ only when the corresponding integral can be represented as a double finite sum \cite{LopezRosa1,Aptekarev1} and it was shown, in particular, that for the Rydberg (i.e., high-lying) orbitals the entropies become level-independent ones. However, a full solution that includes, in particular, the detailed analysis of the corresponding uncertainty relations is still missing. One of the main aims of the present research is to close this gap. A convergence test for improper integrals reveals that the R\'{e}nyi and Tsallis  momentum entropies do exist only for the coefficient  $\alpha$ being greater than its threshold value of one quarter at which they diverge. Exact numerical calculations, which are supported at some particular entropy parameters by their analytic formulae, show that the momentum R\'{e}nyi entropy is a decreasing function of $\alpha$, as it follows from the general properties, and its $\alpha\rightarrow\infty$ asymptotics is explicitly written for the lowest state. At $\alpha=1/2$, the ground orbital turns both uncertainty relations into the equality, as it was noticed before for other structures \cite{Olendski1}. Concrete numerical examples confirm the earlier analytic claim \cite{LopezRosa1} of the independence of the momentum components on the index $n$ at large values of the latter. Since a comparison between different boundary conditions (BCs) imposed on enclosed systems is a common standard practice in quantum chemistry \cite{Fernandez1,Fernandez2}, a research below is also extended to the Neumann well (for which a derivative of the function vanishes at the edges) and it points out similarities and differences between the entropies behavior for the two structures; it is confirmed, in particular, that the BCs governing the {\em position} wave function play a crucial role in determining a semi infinite range of the entropy parameter $\alpha$ where {\em momentum} components exist: for the latter edge requirements the critical value of the R\'{e}nyi or Tsallis factor is one half. A study of the Neumann quantum well is of a huge fundamental significance since it is known that its ground orbital does violate Heisenberg uncertainty $\Delta x\Delta k\geq1/2$ whereas Shannon and R\'{e}nyi relations hold true \cite{Bialynicki3,Bialynicki4}. Our analysis of relations~\eqref{TsallisInequality1} and \eqref{RenyiUncertainty2} reveals that, similar to all other studied before magnetic-field-free structures \cite{Olendski1}, they for the lowest-energy state transform at $\alpha=1/2$ into the equalities with their either side being equal to $\ln(2\pi)$ (R\'{e}nyi functionals) or $\Phi_1(0)$ (Tsallis measures). It is shown also that the mismatch in the threshold parameters causes different behavior of the R\'{e}nyi sum from Equation~\eqref{RenyiUncertainty2} at large $\alpha$: it tends to the finite limit for the Dirichlet BC and unrestrictedly increases for the Neumann requirement what results in the unconstrained loss of the overall information about the latter structure. The only previous discussion on the dependence of the R\'{e}nyi and Tsallis entropies on non-Dirichlet BCs addressed their behavior for the sole orbital of the attractive Robin wall \cite{Olendski1}. In this way, our comparative analysis enriches very scarce knowledge  in this field; in particular, summing up the results of these two endeavors, one can conjecture that $R_\gamma(\alpha)$ and $T_\gamma(\alpha)$ for any non-Dirichlet system do exist at $\alpha\geq1/2$ only. Of course, this surmise has to be further investigated. From the point of view of quantum information processing, a significance of the present research lies in the fact that the Dirichlet and Neumann edge conditions describe different types of materials that can be used for the design of devices for data compression, quantum cryptography, entanglement witnessing, quantum metrology and other tasks employing entropic uncertainty relations \cite{Coles1,Wang1}: position waveform vanishing at the surface is a good approximation for semiconductor nanostructures and the Neumann requirement is relevant for the analysis of superconductors \cite{deGennes1}. Accordingly, depending on the task to be performed, one of these substances might present a better choice.

\section{Dirichlet well}
\begin{figure}[bt]
\centering
\includegraphics[width=\columnwidth]{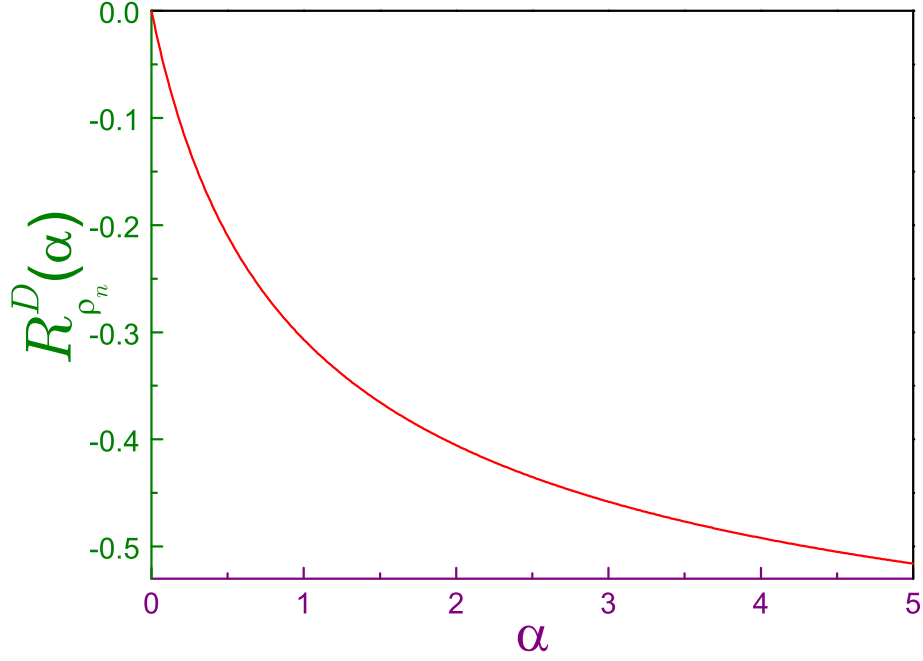}
\caption{\label{DirichletRenyiPositionFig1}R\'{e}nyi position entropy of the Dirichlet well as a function of the parameter $\alpha$. The well width is assumed to be equal to unity, $a\equiv1$.}
\end{figure}

Consider a 1D quantum particle with mass $m^*$ that is free to move inside the interval $-a/2<x<a/2$ and at the edges its position waveform $\Psi^D(x)$ vanishes:
\begin{equation}\label{Dirichlet2}
\Psi^D(-a/2)=\Psi^D(a/2)=0.
\end{equation}
Its discrete energy spectrum reads:
\begin{equation}\label{DirichletEnergy1}
E_n^D=\frac{\pi^2\hbar^2}{2m^*a^2}n^2,\quad n=1,2,\ldots,
\end{equation}
the corresponding eigen waveforms of the Schr\"{o}dinger equation are expressed with the help of trigonometric functions:
\begin{subequations}\label{DirichletFunc1}
\begin{align}\label{DirichletFuncPosition1}
\Psi_n^D(x)&=-\left(\frac{2}{a}\right)^{1/2}\sin\frac{n\pi}{a}\!\left(\!x-\frac{a}{2}\right),
\intertext{whereas their momentum counterparts are:}
\label{DirichletFuncMomentum1}
\Phi_n^D(k)&=\left(\frac{a}{\pi}\right)^{1/2}\frac{n\pi\left[1-(-1)^ne^{-iak}\right]}{(n\pi)^2-(ak)^2}e^{-iak/2}.
\end{align}
\end{subequations}
They are orthonormalized according to Equation~\eqref{OrthoNormality1}:
\begin{equation}\label{OrthoNorm2}
\int_{-a/2}^{a/2}\Psi_{n'}(x)\Psi_n(x)dx=\int_{-\infty}^\infty\Phi_{n'}^*(k)\Phi_n(k)dk=\delta_{nn'}.
\end{equation}
Corresponding densities
\begin{subequations}\label{DirichletDensities1}
\begin{align}\label{DirichletPositionDensity1}
\rho_n^D(x)&=\frac{2}{a}\sin^2\frac{\pi n}{a}\!\left(\!x-\frac{a}{2}\right)\\
\label{DirichletMomentumDensity1}
\gamma_n^D(k)&=\frac{4a}{\pi}\left[\frac{n\pi}{(ak)^2-(n\pi)^2}\sin\frac{ak-n\pi}{2}\right]^2
\end{align}
\end{subequations}
define the R\'{e}nyi entropies as:
\begin{subequations}\label{DirichletRenyi1}
\begin{align}\label{DirichletRenyiPosition1}
R_{\rho_n}^D(\alpha)&=\ln a +\frac{1}{1-\alpha}\ln\!\left(\frac{2^\alpha}{\pi}\int_0^\pi(\sin^2z)^\alpha dz\right)\\
\label{DirichletRenyiMomentum1}
R_{\gamma_n}^D(\alpha)&=-\ln a+\ln2 +\frac{1}{1-\alpha}\!\ln\!\!\left((n^2\pi)^\alpha\!\!\!\int_0^\infty\left(\left[\frac{\sin\left(z-n\pi/2\right)}{z^2-\left(n\pi/2\right)^2}\right]^2\right)^\alpha\!\!dz\!\!\right).
\end{align}
\end{subequations}
These equations manifest that the position (momentum) component depends on the well spatial extent $a$ as a positive (negative) logarithm of this characteristic distance of the system what results in the width independent sum entering, e.g., the uncertainty relation, Equation~\eqref{RenyiUncertainty2}. Next, it immediately follows from Equation~\eqref{DirichletRenyiPosition1} that the position R\'{e}nyi entropy $R_n^D(\alpha)$ has the same variation for all orbitals and is defined for any non negative coefficient $\alpha$. The integral in Equation~\eqref{DirichletRenyiPosition1} can be evaluated analytically \cite{Gradshteyn1,Prudnikov1} yielding
\begin{equation}\tag{23a$'$}\label{eq:23a'}
R_{\rho_n}^D(\alpha)=\ln a+\frac{1}{1-\alpha}\ln\!\left(\frac{2^\alpha}{\pi^{1/2}}\frac{\Gamma\left(\alpha+\frac{1}{2}\right)}{\Gamma(\alpha+1)}\right),
\end{equation}
as pointed out first by J. S\'{a}nchez-Ruiz \cite{SanchezRuiz1}. Integer values $\alpha\geq2$ \cite{SanchezRuiz1,Aptekarev1} simplify this expression to
\begin{subequations}\label{DirichletRenyiAsymptote1}
\begin{align}\label{DirichletRenyiAsymptote1_Integer}
R_{\rho_n}^D(m)&=\ln a+\frac{1}{1-m}\ln\frac{(2m)!}{2^m(m!)^2},\quad m=2,3,\ldots;
\intertext{at the vanishing R\'{e}nyi parameter, the position component approaches $\ln a$ from below as}
\label{DirichletRenyiAsymptote1_0}
R_{\rho_n}^D(\alpha)&=\ln a -(\ln2)\alpha+\left(\frac{\pi^2}{6}-\ln2\right)\alpha^2+\ldots,\quad\alpha\rightarrow0;
\intertext{near $\alpha=1/2$ it behaves as:}
\label{DirichletRenyiAsymptote1_OneHalf}
R_{\rho_n}^D(\alpha)&=\ln a+\ln\frac{8}{\pi^2}+4\left(\ln\frac{8}{\pi}-1\right)\left(\alpha-\frac{1}{2}\right)+\ldots,\quad\alpha\rightarrow\frac{1}{2},
\intertext{and it turns into its Shannon counterpart according to}
\label{DirichletRenyiAsymptote1_1}
R_{\rho_n}^D(\alpha)&=\ln a-1+\ln2+\frac{1}{2}\left(3-\frac{\pi^2}{3}\right)(\alpha-1)+\ldots,\quad\alpha\rightarrow1,
\intertext{whereas for large $\alpha$ its variation is:}
\label{DirichletRenyiAsymptote1_Infinite}
R_{\rho_n}^D(\alpha)&=\ln a -\ln2+\frac{1}{\alpha}\ln\frac{\pi^{1/2}\alpha^{1/2}}{2}+\ldots,\quad\alpha\rightarrow\infty.
\end{align}
\end{subequations}
The numbers at the first nonvanishing expansion coefficients in Equations~\eqref{DirichletRenyiAsymptote1_OneHalf} and \eqref{DirichletRenyiAsymptote1_1} are negative:
$$4\left(\ln\frac{8}{\pi}-1\right)=-0.2611\ldots,\quad\frac{1}{2}\left(3-\frac{\pi^2}{3}\right)=-0.1449\ldots,$$
what is consistent with the general property of the decrease of the R\'{e}nyi entropy with the increasing factor $\alpha$.
The leading terms in Equations~\eqref{DirichletRenyiAsymptote1_OneHalf}--\eqref{DirichletRenyiAsymptote1_Infinite} lie below zero too:
$$\ln\frac{8}{\pi^2}=-0.2100\ldots,\quad-1+\ln2=-0.3068\ldots,\quad-\ln2=-0.6931\ldots,$$ 
what means that the position entropy [or, more precisely, the expression $R_{\rho_n}^D(\alpha)-\ln a$] at the positive R\'{e}nyi parameter is negative and turns to zero together with $\alpha$. This is reflected in Figure~\ref{DirichletRenyiPositionFig1} exhibiting a dependence of the quantity $R_{\rho_n}^D(\alpha)-\ln a$ on the R\'{e}nyi coefficient. It is a monotonically decreasing function of $\alpha$ that vanishes together with it by the law from Equation~\eqref{DirichletRenyiAsymptote1_0} and approaches $-\ln2$ at the large parameter.

\begin{figure}[bt]
\centering
\includegraphics[width=\columnwidth]{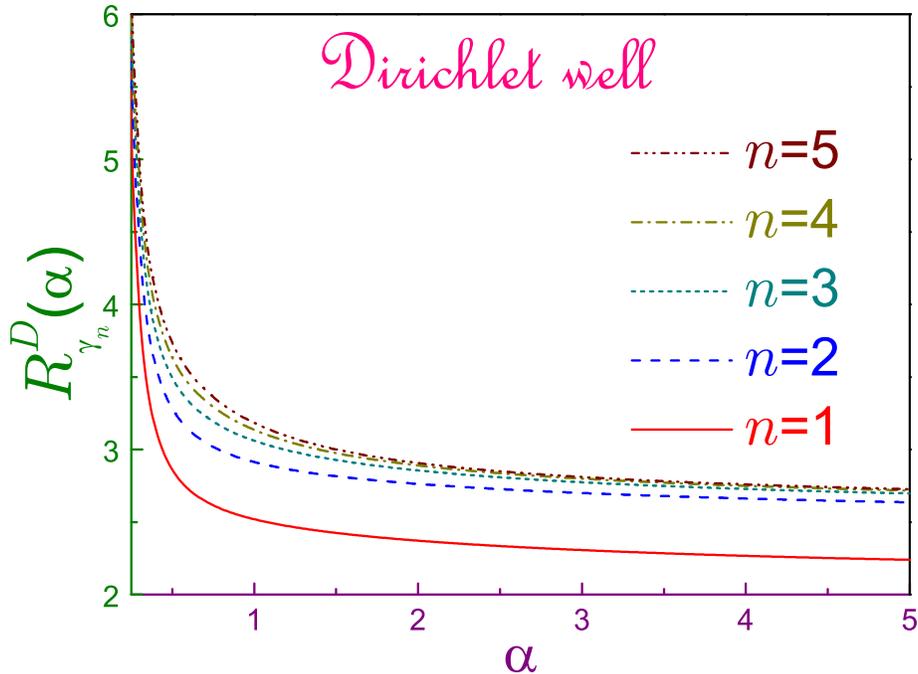}
\caption{\label{DirichletRenyiMomentumFig1}R\'{e}nyi momentum entropies $R_{\gamma_n}^D(\alpha)$  of the Dirichlet unit-width well as functions of the parameter $\alpha$ where solid line depicts the functional of the ground orbital, dashed curve is for the first excited state, dotted one -- for the level with $n=3$, dash-dotted dependence is for the state with $n=4$, and dash-dot-dotted one - for $n=5$.}
\end{figure}

At the integer $\alpha$, an improper integral entering Equation~\eqref{DirichletRenyiMomentum1} can be calculated as a double finite sum \cite{LopezRosa1,Aptekarev1}. For the arbitrary real R\'{e}nyi coefficient, applying to it a comparison convergence test \cite{Taylor1}, one concludes that there is a level-independent bound $\alpha_{TH}^D$ that limits from below a semi-infinite range $\left[\alpha_{TH}^D,+\infty\right)$ at which the Dirichlet momentum entropy $R_{\gamma_n}^D(\alpha)$ does exist:
\begin{equation}\label{DirichletThreshold1}
\alpha_{TH}^D=\frac{1}{4}.
\end{equation}
Note that the quasi-1D hydrogen atom, which is another Dirichlet structure, is characterized by the same critical value \cite{Olendski1}. As the R\'{e}nyi parameter approaches this threshold, the corresponding entropy does diverge, as it follows from its general properties and is depicted in Figure~\ref{DirichletRenyiMomentumFig1} that shows $R_{\gamma_n}^D(\alpha)+\ln a$ behavior for the five lowest states. For some values of $\alpha$, momentum R\'{e}nyi components can be evaluated analytically; for example, at $\alpha=1/2$, which will be used in the discussion of the uncertainty relation, one has for the two lowest states:
\begin{subequations}\label{DirichletRenyiAsymptoteMomentumOneHalf1}
\begin{align}\label{DirichletRenyiAsymptoteMomentumOneHalf1_1}
R_{\gamma_1}^D\left(\frac{1}{2}\right)&=-\ln a+2\ln\!\left(\frac{4}{\pi^{1/2}}{\rm Si}(\pi)\right)\\
\label{DirichletRenyiAsymptoteMomentumOneHalf1_2}
R_{\gamma_2}^D\left(\frac{1}{2}\right)&=-\ln a+2\ln\!\left(\frac{4}{\pi^{1/2}}\left[2{\rm Si}(\pi)-{\rm Si}(2\pi)\right]\right),
\end{align}
\end{subequations}
where ${\rm Si}(z)$ is sine integral \cite{Abramowitz1} and numerical values of the last items in the right-hand sides of Equations~\eqref{DirichletRenyiAsymptoteMomentumOneHalf1_1} and \eqref{DirichletRenyiAsymptoteMomentumOneHalf1_2} are $2.8603\ldots$ and $3.2812\ldots$, respectively. Calculations show that at any R\'{e}nyi coefficient the momentum entropy is an increasing function of the quantum index and, as expected, decreases as $\alpha$ grows staying, however, always positive. Our exact numerical computations substantiate an analytic conclusion of the entropy independence on the orbital at large $n$ \cite{LopezRosa1}; namely, as Figure~\ref{DirichletRenyiMomentumFig1} demonstrates, the clustering of $R_{\gamma_n}(\alpha)$ starts to form already at quite small index, $n\sim3-5$, and initially it takes place at the large parameter $\alpha$. For the higher lying levels, the range of the latter where the momentum components are almost indistinguishable from each other, expands to the left. However, the lowest state momentum entropy is split off from its $n\geq2$ counterparts at any R\'{e}nyi parameter. This has crucial consequences for the corresponding uncertainty relations.

\begin{figure}[bt]
\centering
\includegraphics[width=\columnwidth]{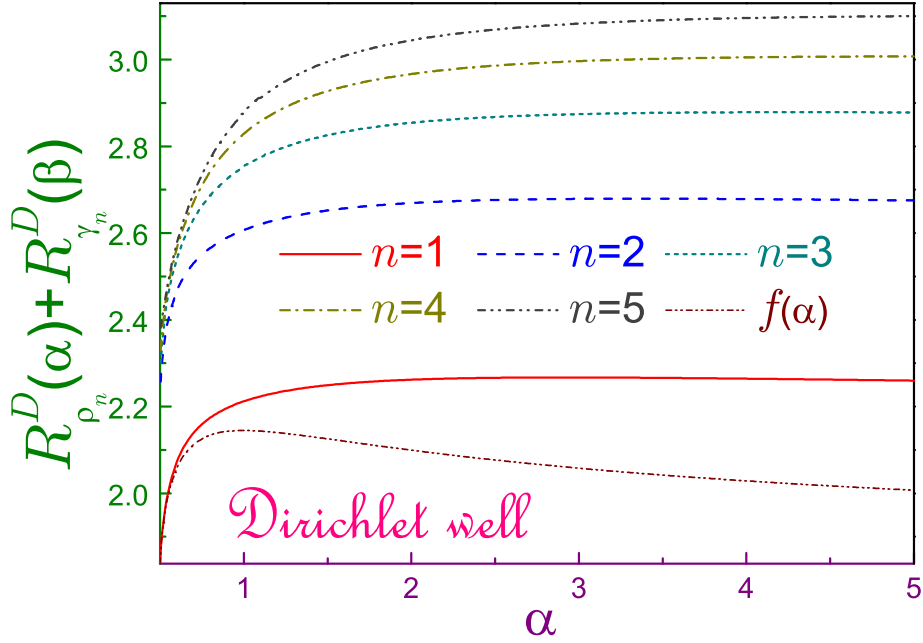}
\caption{\label{DirichletRenyiUncertaintyFig1}Sum of the position and momentum  R\'{e}nyi entropies $R_{\rho_n}^D(\alpha)+R_{\gamma_n}^D(\beta)$ of the Dirichlet well as functions of the parameter $\alpha$ where solid line depicts the sum of the functionals of the ground orbital, dashed curve is for the first excited state, dotted one -- for the level with $n=3$, dash-dotted dependence is for the state with $n=4$, upper (thick) dash-dot-dotted line is for $n=5$ and  its lower (thin) counterpart depicts function $f(\alpha)$ from Equation~\eqref{Function_f1}.}
\end{figure}

Figure~\ref{DirichletRenyiUncertaintyFig1} shows left-hand sides of the R\'{e}nyi inequality, Equation~\eqref{RenyiUncertainty2}, for the five lowest-energy states together with its right-hand side represented by the function \cite{Olendski1}
\begin{equation}\label{Function_f1}
f(\alpha)=\ln\pi-\left[\ln\alpha-\frac{\alpha-1/2}{\alpha-1}\ln(2\alpha-1)\right],
\end{equation}
in terms of the coefficient $\alpha$. A remarkable property of these dependencies is the saturation of the uncertainty relation for the ground state at the left edge of the interval $\left[1/2,+\infty\right)$ when either side of Equation~\eqref{RenyiUncertainty2} turns to $\ln2\pi=1.8378\ldots$, in accordance with the previous observations for other structures \cite{Olendski1}. As the position component at this R\'{e}nyi coefficient is equal to $\ln a+\ln\left(8/\pi^2\right)$, Equation~\eqref{DirichletRenyiAsymptote1_OneHalf}, and the conjugate parameter $\beta$ takes unrestrictedly high values at $\alpha\rightarrow1/2$, one immediately deduces the infinite asymptote of the lowest-orbital momentum entropy:
\begin{equation}\label{DirichletAsymptote2_MomentumInfinity}
R_{\gamma_1}^D(\infty)=-\ln a+\ln\frac{\pi^3}{4}=-\ln\gamma_1^D(0),
\end{equation}
with $\ln\left(\pi^3/4\right)=2.0478\ldots$. For any other R\'{e}nyi parameter, the ground-level uncertainty relation takes a form of a strict inequality as it does for any excited state in the whole range $1/2\leq\alpha<+\infty$. The sum $R_{\rho_n}^D(\alpha)+R_{\gamma_n}^D(\beta)$ increases with the quantum index and as a function of $\alpha$ it exhibits a very broad $n$-dependent maximum; for example, for the lowest (first excited) orbital a maximum of $2.2670\ldots$ ($2.6793\ldots$) is achieved at $\alpha_{{\rm max}_1}\approx2.92$ ($\alpha_{{\rm max}_2}\approx3.35$). After the extremum, the sums slowly decrease to their asymptotic values, which for the two lowest states, by virtue of Equations~\eqref{DirichletRenyiAsymptote1_Infinite} and \eqref{DirichletRenyiAsymptoteMomentumOneHalf1}, are:
\begin{subequations}\label{DirichletRenyiUncertaintyAsymptote1}
\begin{align}\label{DirichletRenyiUncertaintyAsymptote1_1}
\left.R_{\rho_1}^D(\alpha)+R_{\gamma_1}^D(\beta)\right|_{\alpha=\infty}&=\ln\!\left(\frac{8}{\pi}{\rm Si}^2(\pi)\right)=2.1671\ldots\\
\label{DirichletRenyiUncertaintyAsymptote1_2}
\left.R_{\rho_2}^D(\alpha)+R_{\gamma_2}^D(\beta)\right|_{\alpha=\infty}&=\ln\!\left(\frac{8}{\pi}\left[2{\rm Si}(\pi)-{\rm Si}(2\pi)\right]^2\right)=2.5880\ldots.
\end{align}
\end{subequations}

Returning to Equation~\eqref{DirichletAsymptote2_MomentumInfinity}, one remarks that the second equality there appeared directly from comparison with Equation~\eqref{DirichletMomentumDensity1}, which also shows that the ground-state momentum density reaches its global maximum just at the zero momentum, $k=0$. Thus, combining the first and last equalities in Equation~\eqref{DirichletAsymptote2_MomentumInfinity}, one arrives at the continuous distribution generalization of the result known for the discrete random events. Note that for the excited states the largest density maxima $\gamma_{n_{max}}^D$ are achieved at the non zero $n$-dependent wave vectors $k_{n_{max}}^D$, which for the high-lying orbitals read:
\begin{subequations}\label{DirichletMomentumMaxima1}
\begin{align}\label{DirichletMomentumMaxima1_K}
k_{n_{max}}^D=\pm\frac{n\pi}{a}\left(1-\frac{6}{n^2\pi^2}\right),\quad n\rightarrow\infty,
\intertext{that corresponds to}
\label{DirichletMomentumMaxima1_G}
\gamma_{n_{max}}^D=\frac{a}{4\pi}\left(1+\frac{3}{n^2\pi^2}\right),\quad n\rightarrow\infty.
\intertext{Last equation shows that for the Rydberg states the maximum becomes practically a level-independent quantity. Plots of the functions $\gamma_n^D(k)$ for the index $n$ up to 11 are presented in Refs.~\cite{Robinett1,Majernik1,Majernik2,Olendski2}. Relation~\eqref{DirichletMomentumMaxima1_G} helps, in turn, to find the momentum R\'{e}nyi entropies at large $\alpha$:}
\label{DirichletMomentumRenyiInfinity1}
R_{\gamma_n}^D(\infty)=-\ln a +\ln(4\pi)-\frac{3}{n^2\pi^2},\quad n\rightarrow\infty,
\end{align}
\end{subequations}
with $\ln(4\pi)=2.5310\ldots$. This supplements an earlier conclusion of the independence of the momentum R\'{e}nyi entropy on the quantum index at the large $n$ \cite{LopezRosa1}.
\begin{figure}[bt]
\centering
\includegraphics[width=\columnwidth]{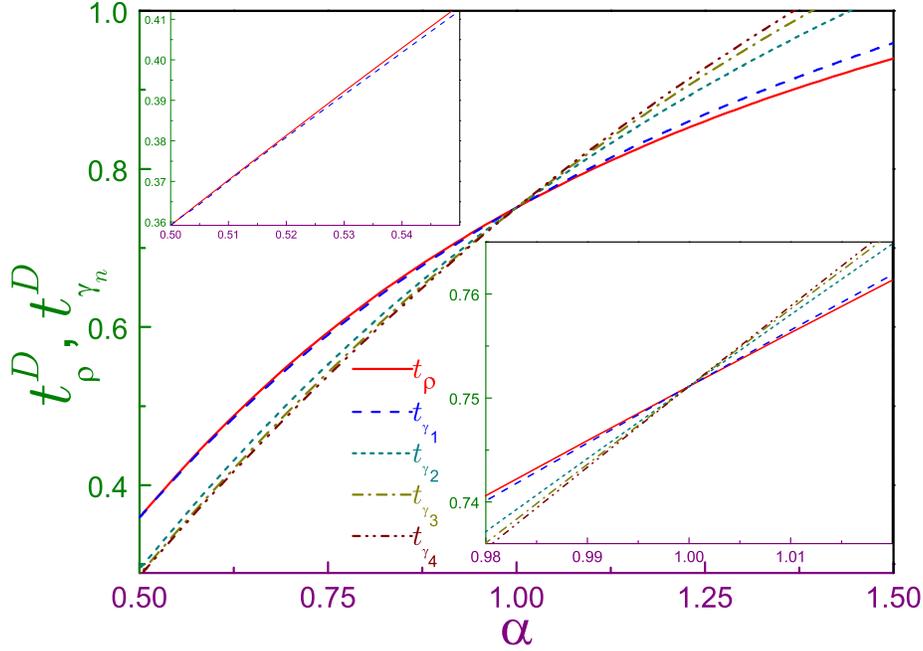}
\caption{\label{DirichletTsallisUncertaintyFig1} Solid line depicts the quantity $t_\rho^D(\alpha)$, Equation~\eqref{DirichletTsallisT1_r}, whereas all other curves correspond to $t_{\gamma_n}^D(\beta)$, Equation~\eqref{DirichletTsallisT1_g}, with the dashed marks being for $n=1$, dotted one, $n=2$, dash-dotted designation is for $n=3$, and dash-dot-dotted line - for $n=4$. Lower right inset shows an enlarged view at the Tsallis parameter close to unity, $\alpha\simeq1$, and upper left window details the behavior of $t_\rho^D(\alpha)$ and $t_{\gamma_1}^D(\beta)$ near $\alpha=1/2$. Factors $\alpha$ and $\beta$ are conjugated according to Equation~\eqref{RenyiUncertainty1}.}
\end{figure}

For the continuous probability distributions, a dimensional incompatibility of the items entering the Tsallis entropies precludes their direct usage  but one can analyze the corresponding uncertainty relation from Equation~\eqref{TsallisInequality1}, or, equivalently, Equation~\eqref{Sobolev1}, which for our geometry read:
\begin{align}
&2^{1/2}a^\frac{1-\alpha}{2\alpha}\alpha^{1/(4\alpha)}\left[\frac{\Gamma\left(\alpha+\frac{1}{2}\right)}{\pi\Gamma(\alpha+1)}\right]^{1/(2\alpha)}\geq\nonumber\\
\label{DirichletTsallisUncertainty1}
&\geq2^\frac{1-\beta}{\beta}a^\frac{\beta-1}{2\beta}n\pi^\frac{2\beta-1}{4\beta}\beta^{1/(4\beta)}\left[\int_0^\infty\!\!\left(\!\left[\frac{\sin\!\left(z-\frac{n\pi}{2}\right)}{z^2-\frac{n^2\pi^2}{4}}\right]^2\right)^{\!\!\beta}\!\!dz\right]^{1/(2\beta)}.
\end{align}
Note that due to the conjugation requirement from Equation~\eqref{RenyiUncertainty1}, above inequality is dimensionally correct. Observe also that its left-hand side is a level-independent one.

Figure~\ref{DirichletTsallisUncertaintyFig1} depicts dimensionless left- and right-hand sides of Equation~\eqref{DirichletTsallisUncertainty1}, i.e., the quantities
\begin{subequations}\label{DirichletTsallisT1}
\begin{align}\label{DirichletTsallisT1_r}
t_\rho^D(\alpha)&=2^{1/2}\alpha^{1/(4\alpha)}\left[\frac{\Gamma\left(\alpha+\frac{1}{2}\right)}{\pi\Gamma(\alpha+1)}\right]^{1/(2\alpha)}\\
\label{DirichletTsallisT1_g}
t_{\gamma_n}^D(\beta)&=2^\frac{1-\beta}{\beta}n\pi^\frac{2\beta-1}{4\beta}\beta^{1/(4\beta)}\left[\int_0^\infty\!\!\left(\!\left[\frac{\sin\!\left(z-\frac{n\pi}{2}\right)}{z^2-\frac{n^2\pi^2}{4}}\right]^2\right)^{\!\!\beta}\!\!dz\right]^{1/(2\beta)},
\end{align}
\end{subequations}
as functions of the Tsallis coefficent $\alpha$. As expected, the entropic inequality, Equation~\eqref{TsallisInequality1},
holds true inside the interval from Equation~\eqref{Sobolev2} only and is saturated by all orbitals at $\alpha=1$. At the arbitrary Tsallis factor, the difference between $t_\rho^D$ and $t_{\gamma_n}^D$ is the smallest for the ground level, $n=1$. In fact, as Figure~\ref{DirichletTsallisUncertaintyFig1} demonstrates, the momentum components of the Tsallis entropies of the excited states are practically indistinguishable from each other already at $n\geq3$. Explanation of this phenomenon is the same as for the R\'{e}nyi functionals \cite{LopezRosa1}. Similar to them, the lowest-energy quantity $t_{\gamma_1}^D$ is split off from its counterparts and, as upper left inset demonstrates, tightens the entropic relation at $\alpha=1/2$. To explain this saturation, one has to point out that, at this value of the Tsallis factor, the left-hand side of Equation~\eqref{Sobolev1} for our geometry turns to
\begin{equation}\label{TsallisCase1}
\frac{1}{(2\pi)^{1/2}}\int_{-a/2}^{a/2}\left|\Psi_n(x)\right|dx,
\end{equation}
what for the ground orbital, $n=1$, when the corresponding waveform $\Psi_1(x)$ does not change sign, becomes $\Phi_1(0)$,
cf. Equation~\eqref{Fourier1_1}, and this value is equal, according to Equation~\eqref{DirichletFuncMomentum1}, to
\begin{equation}\label{RelationDirichlet1}
\Phi_1^D(0)=\frac{2a^{1/2}}{\pi^{3/2}}=a^{1/2}\,0.3591\ldots.
\end{equation}
It is easily shown that in the neighborhood of $\alpha=1/2$ the left-hand side of Equation~\eqref{DirichletTsallisUncertainty1}, which, as mentioned above, is valid accidentally for any state, behaves as
\begin{equation}\label{TsallisCase3}
\Phi_1^D(0)\left[1+\left(\ln\frac{2\pi^3}{a^2}-1\right)\left(\alpha-\frac{1}{2}\right)+\ldots\right],\quad\alpha\rightarrow\frac{1}{2}.
\end{equation}
On the other hand, the same limit, which corresponds to the infinite values of the conjugated factor, $\beta=\infty$, reduces the right-hand side of the Sobolev relation to 
\begin{equation}\label{TsallisCase2}
\left|\Phi_n(k)\right|_{max}.
\end{equation}
As discussed above, for the ground level, $n=1$, this maximum is achieved just at the zero momentum. Thus, we have proved that for the lowest-energy state the left- and right-hand sides of Equation~\eqref{DirichletTsallisUncertainty1} at $\alpha=1/2$ are equal to each other and are $\Phi_1(0)$.

\section{Neumann well}
For this BC, which is written as
\begin{equation}\label{NeumannBC1}
\left.\frac{d\Psi_n^N(x)}{dx}\right|_{x=\mp a/2}=0,
\end{equation}
the waveforms are:

in the position space
\begin{subequations}\label{NeumannFunc1}
\begin{align}\label{NeumanntFuncPosition1}
\Psi_n^N(x)&=\left\{\begin{array}{cc}
a^{-1/2},&n=1\\
\left(\frac{2}{a}\right)^{1/2}\cos\frac{(n-1)\pi}{a}\!\left(x-\frac{a}{2}\right),&n\geq2\end{array}\right.;\\
\intertext{in the wave vector representation \cite{Olendski2}:}
\label{NeumannFuncMomentum1}
\Phi_n^N(k)&=\left\{\begin{array}{cc}
\left(\frac{a}{2\pi}\right)^{1/2}\frac{2}{ak}\sin\frac{ak}{2},&n=1\\
-\left(\frac{a}{\pi}\right)^{1/2}\frac{iak\left[1+(-1)^ne^{-iak}\right]}{[(n-1)\pi]^2-(ak)^2}e^{-iak/2},&n\geq2\end{array}\right..
\end{align}
\end{subequations}
It is important to underline here that, as mentioned in the Introduction, Equation~\eqref{Domain1}, the Neumann position function is {\em not} defined outside the well; in particular, it is {\em not} zero at $|x|>a/2$ for, if it were the case, the discontinuity at the boundaries will lead to the infinite momentum $-i\hbar d/dx$ at $x=\pm a/2$. Physically, this BC corresponds to, e.g., superconducting film \cite{deGennes1} where the order parameter $\Psi$ exists only inside the zero-resistance material. The associated energy spectrum
\begin{equation}\label{NeumannEnergy1}
E_n^N=\frac{\pi^2\hbar^2}{2m^*a^2}(n-1)^2,\quad n=1,2,\ldots,
\end{equation}
differs from its Dirichlet counterpart, Equation~\eqref{DirichletEnergy1}, by the presence of the zero energy state, $E_1^N=0$, whose position wave function is just a constant, as it follows from Equation~\eqref{NeumanntFuncPosition1}. Expressions for the densities \cite{Olendski2}
\begin{subequations}\label{NeumannDensities1}
\begin{align}\label{NeumannDensityPosition1}
\rho_n^N(x)&=\left\{\begin{array}{cc}
1/a,&n=1\\
\frac{2}{a}\cos^2\frac{(n-1)\pi}{a}\!\left(x-\frac{a}{2}\right),&n\geq2\end{array}\right.;\\
\label{NeumannDensitiesMomentum1}
\gamma_n^N(k)&=\left\{\begin{array}{cc}
\frac{a}{2\pi}\left(\frac{2}{ak}\sin\frac{ak}{2}\right)^2,&n=1\\
\frac{4a}{\pi}\left[\frac{ak}{(ak)^2-[(n-1)\pi]^2}\sin\frac{ak-(n-1)\pi}{2}\right]^2,&n\geq2\end{array}\right.\end{align}
\end{subequations}
lead to the R\'{e}nyi entropies:
\begin{subequations}\label{NeumannRenyi1}
\begin{align}\label{NeumannRenyiPosition1}
R_{\rho_n}^N(\alpha)&=\left\{\begin{array}{cc}
\ln a,&n=1\\
\ln a +\frac{1}{1-\alpha}\ln\left(\frac{2^\alpha}{\pi}\int_0^\pi(\cos^2z)^\alpha dz\right),&n\geq2\end{array}\right.;\\
\label{NeumannRenyiMomentum1}
R_{\gamma_n}^N(\alpha)&=\left\{\begin{array}{cc}
-\ln a+\frac{1}{1-\alpha}\ln\left(\frac{2^{-\alpha+2}}{\pi^\alpha}\int_0^\infty\left[\left(\frac{\sin z}{z}\right)^2\right]^\alpha dz\right),&n=1\\
-\ln a +\frac{1}{1-\alpha}\ln\left(\frac{4}{\pi^\alpha}\int_0^\infty\left(\left[\frac{z\sin\left(z-n'\pi/2\right)}{z^2-\left(n'\pi/2\right)^2}\right]^2\right)^\alpha dz\right),&n\geq2\end{array}\right.,
\end{align}
\end{subequations}
where for brevity a designation $n'=n-1$ has been used. First thing to notice is the fact that the Neumann ground state has a coefficient-independent position entropy $\ln a$ what is due to the constant value of the corresponding waveform. Second, position R\'{e}nyi entropies of the excited orbitals are again the level-independent quantities that are exactly the same as their Dirichlet counterpart and so, the whole discussion of the latter, including Equations~\eqref{eq:23a'} and \eqref{DirichletRenyiAsymptote1} and Figure~\ref{DirichletRenyiPositionFig1}, straightworfardly applies to the Neumann BC too. Third, momentum entropies exist only at the R\'{e}nyi or Tsallis parameter that is greater than 
\begin{equation}\label{NeumannThreshold1}
\alpha_{TH}^N=\frac{1}{2},
\end{equation}
which is twice of its Dirichlet counterpart, Equation~\eqref{DirichletThreshold1}. Observe that the same critical coefficient was found for the attractive Robin wall \cite{Olendski1} what might lead to the conjecture that any non-Dirichlet BC for the 1D system has just this magnitude of $\alpha_{TH}$. To support this claim even more, let us point out that the form of the momentum density of the mixed Dirichlet-Neumann BCs \cite{Olendski2} leads to the same threshold, as for the pure Neumann structure, Equation~\eqref{NeumannThreshold1}. Next, at the integer $\alpha$ the integral entering ground-state momentum entropy is evaluated analytically \cite{Prudnikov1}:
\begin{equation}\label{Integral1}
\int_0^\infty\left(\frac{\sin z}{z}\right)^{2m}dz=m\pi\sum_{j=0}^{m-1}\frac{(-1)^j}{j!(2m-j)!}(m-j)^{2m-1},\quad m=1,2,\ldots.
\end{equation}
Note that this result can be also obtained as a particular case of Lemma 3 in Reference~\cite{Aptekarev1}. In addition, lowest energy wave vector R\'{e}nyi functional approaches the Shannon value as
\begin{subequations}\label{LimitsRenyiNeumann1}
\begin{align}\label{LimitsRenyiNeumann1_1}
R_{\gamma_1}^N(\alpha)&=-\ln a+\ln(2\pi)+2(1-\gamma)+2(\gamma-1)^2(\alpha-1)+\ldots,\quad\alpha\rightarrow1,
\intertext{where $\gamma=0.5772\ldots$ is Euler's constant \cite{Abramowitz1}, and its value at infinity is}
\label{LimitsRenyiNeumann1_Infinity}
R_{\gamma_1}^N(\infty)&=-\ln a+\ln(2\pi).
\end{align}
\end{subequations}
In deriving Equation~\eqref{LimitsRenyiNeumann1_1},  the value of the integral \cite{SanchezRuiz2}
\begin{equation}\label{Integral2}
\int_0^\infty\frac{\sin^2z}{z^2}\ln\!\left(\frac{\sin^2z}{z^2}\right)dz=-\pi(1-\gamma)=-1.3282\ldots
\end{equation}
has been used.
\begin{figure}[bt]
\centering
\includegraphics[width=\columnwidth]{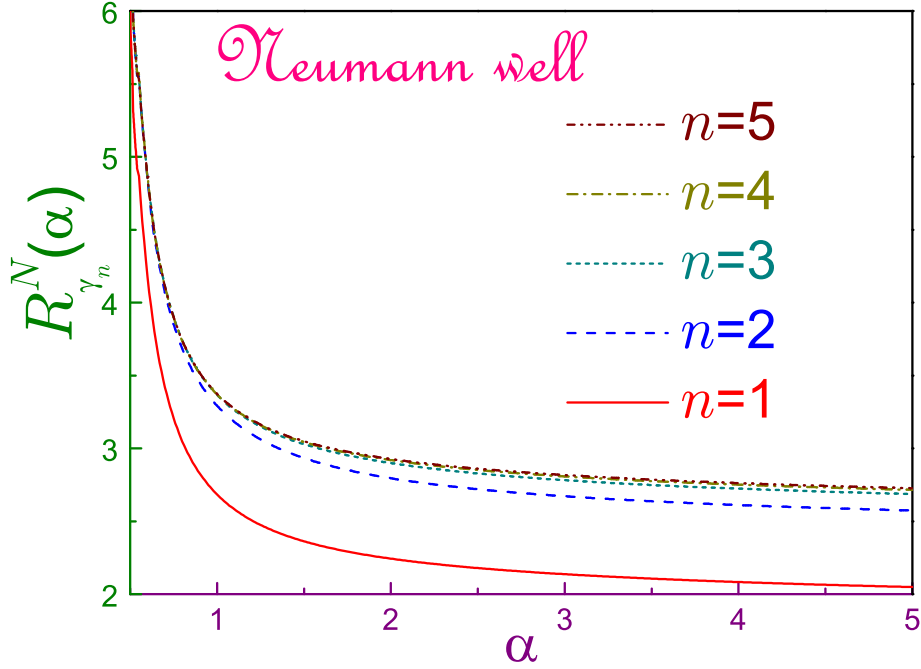}
\caption{\label{NeumannRenyiMomentumFig1}R\'{e}nyi momentum entropies $R_{\gamma_n}^N(\alpha)$  of the Neumann unit-width well as functions of the parameter $\alpha$. The same convention as in Figure~\ref{DirichletRenyiMomentumFig1} is used.}
\end{figure}

It is seen from Figure~\ref{NeumannRenyiMomentumFig1}, which depicts momentum R\'{e}nyi entropies of the Neumann orbitals, that, as expected, they logarithmically diverge at the coefficient $\alpha$ approaching its critical magnitude from Equation~\eqref{NeumannThreshold1}. Similar to the Dirichlet configuration, all entropies grow with the index $n$, stay positive at any R\'{e}nyi parameter and at the large $\alpha$ the lowest-state entropy is split off from its neighbors, which only slightly differ from each other already at $n\geq3$. For the Rydberg orbitals, $n\gg1$, the momentum density largest maxima,
\begin{subequations}\label{NeumannMomentumMaxima1}
\begin{align}\label{NeumannMomentumMaxima1_G}
\gamma_{n_{max}}^N&=\frac{a}{4\pi}\left(1+\frac{8}{3n^2\pi^2}\right),
\intertext{which, similar to their Dirichlet counterparts, Eq.~\eqref{DirichletMomentumMaxima1_G}, are almost level-independent quantities, are achieved at}
\label{NeumannMomentumMaxima1_K}
k_{n_{max}}^N&=\pm\frac{(n-1)\pi}{a}\left(1+\frac{4}{n^2\pi^2}\right),
\intertext{which, as compared to the Dirichlet well, Eq.~\eqref{DirichletMomentumMaxima1_K}, are located by $\pi/a$ closer to the zero wave vector. This difference can be seen in Figure~2 of Reference~\cite{Olendski2} that provides a comparative analysis of the two types of the momentum densities. According to Equation~\eqref{NeumannMomentumMaxima1_G}, the entropies at the infinitely high R\'{e}nyi factor take the form:}
\label{NeumannMomentumRenyiInfinity1}
R_{\gamma_n}^N(\infty)&=-\ln a +\ln(4\pi)-\frac{8}{3n^2\pi^2},\quad n\rightarrow\infty.
\end{align}
\end{subequations}

\begin{figure}[bt]
\centering
\includegraphics[width=\columnwidth]{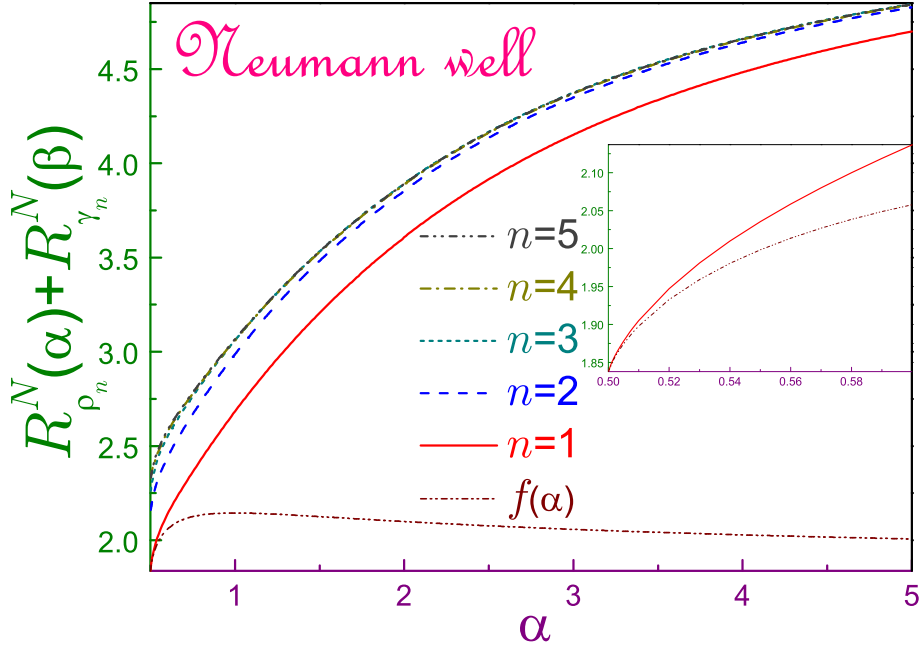}
\caption{\label{NeumannRenyiUncertaintyFig1}Sum of the position and momentum  R\'{e}nyi entropies $R_{\rho_n}^N(\alpha)+R_{\gamma_n}^N(\beta)$ of the Neumann well as functions of the parameter $\alpha$. The same nomenclature as in Figure~\ref{DirichletRenyiUncertaintyFig1} is used. Inset shows an enlarged view of the ground state behavior near the threshold $\alpha_{TH}^N$.}
\end{figure}

From Equations~\eqref{NeumannRenyiPosition1} and \eqref{LimitsRenyiNeumann1_Infinity} it immediately follows that the Neumann ground level does saturate at $\alpha=1/2$ the R\'{e}nyi uncertainty relation. A detailed view of the behavior in the vicinity of this value of the coefficient is presented in inset of Figure~\ref{NeumannRenyiUncertaintyFig1}. As it follows from the main body of this Figure, the sums entering the left-hand side of Equation~\eqref{RenyiUncertainty2}, become indistinguishable from each other in the whole range $\alpha\geq1/2$ already at $n\geq3$ whereas for the Dirichlet structure it was true for the same quantum indices just very close to the left edge of this interval, cf. Figure~\ref{DirichletRenyiUncertaintyFig1}. Another fundamental difference is their behavior at the large factor $\alpha$. As discussed in the previous Section, the Dirichlet sums $R_{\rho_n}^D(\alpha)+R_{\gamma_n}^D(\beta)$ tend at $\alpha\rightarrow\infty$ to the finite values, which loose their $n$ dependence in the Rydberg regime only. The unconstrained growth of their Neumann counterparts in this limit, which is exemplified in Figure~\ref{NeumannRenyiUncertaintyFig1}, is explained by the different values of the thresholds, Equations~\eqref{DirichletThreshold1} and \eqref{NeumannThreshold1}; namely, for the Neumann well the increasing R\'{e}nyi coefficient $\alpha$ pushes its conjugated partner $\beta$ closer and closer to one-half, what is just the critical parameter from Eq.~\eqref{NeumannThreshold1}; hence, the divergence of the momentum item causes the same behavior of the whole sum $R_{\rho_n}^N(\alpha)+R_{\gamma_n}^N(\beta)$. In terms of the information theory, this means the decrease of our overall knowledge of the behavior of the system.

\begin{figure}[bt]
\centering
\includegraphics[width=\columnwidth]{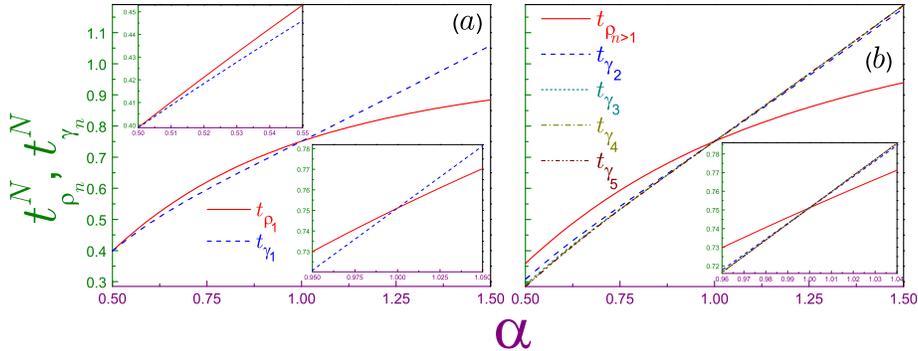}
\caption{\label{NeumannTsallisUncertaintyFig1}Dimensionless parts of the Tsallis uncertainty relations of the Neumann well for a) ground orbital, $n=1$, and b) several excited states where solid curves are for the position components and dashed line in the left panel corresponds to the wave vector contribution whereas its counterpart from the right subplot depicts momentum factor for $n=2$, dotted dependence is for the momentum component with $n=3$, dash-dotted one -- for $n=4$, and dash-dot-dotted curve -- for $n=5$. Insets enlarge the behavior close to the edges of the interval from Equation~\eqref{Sobolev2}.}
\end{figure}

For the ground Neumann orbital, Tsallis entropic relation takes the form:
\begin{subequations}\label{TsallisNeumann1}
\begin{align}\label{TsallisNeumann1_1}
&a^\frac{1-\alpha}{2\alpha}\left(\frac{\alpha}{\pi}\right)^{1/(4\alpha)}\geq\left(\frac{2}{\pi}\right)^{1/2}a^\frac{\beta-1}{2\beta}\left(\frac{\beta}{\pi}\right)^{1/(4\beta)}\left(\int_0^\infty\!\!\left[\!\left(\frac{\sin z}{z}\right)^2\right]^{\!\!\beta}\!\!dz\right)^{1/(2\beta)},
\intertext{which in the neighborhood of $\alpha=1$ degenerates, according to Equation~\eqref{TsallisInequality2}, to}
\label{TsallisNeumann1_2}
&\frac{1}{\pi^{1/4}}\left[1+\left(1+\ln\frac{\pi}{a^2}\right)\frac{\alpha-1}{4}\right]\geq\frac{1}{\pi^{1/4}}\left[1+\left(3-4\gamma+\ln\frac{4\pi}{a^2}\right)\frac{\alpha-1}{4}\right],
\intertext{which is satisfied, as expected, only inside the interval from Equation~\eqref{Sobolev2}, for $3-4\gamma+\ln4=2.0774\ldots$ is greater than unity. At the left edge of this region, the position component behaves as}
\label{TsallisNeumann1_3}
&\Phi_1^N(0)\left[1+\left(1+\ln\frac{2\pi}{a^2}\right)\left(\alpha-\frac{1}{2}\right)\right],\quad\alpha\rightarrow\frac{1}{2},
\end{align}
\end{subequations}
with
\begin{equation}\label{RelationNeumann1}
\Phi_1^N(0)=\left(\frac{a}{2\pi}\right)^{1/2}=a^{1/2}0.3989\ldots,
\end{equation}
as it follows from Equation~\eqref{NeumannFuncMomentum1}. Note that the numerical value of the coefficient in the last equation is greater than its Dirichlet counterpart, Equation~\eqref{RelationDirichlet1}. Evolution of both sides of Equation~\eqref{TsallisNeumann1_1} with the Tsallis parameter $\alpha$ is depicted in panel (a) of Figure~\ref{NeumannTsallisUncertaintyFig1}, which exemplifies that the ground-state position and wave vector parts do coincide at both edges of the interval from Equation~\eqref{Sobolev2}. Figure 7(b) shows dimensionless components of the uncertainty relation for the excited Neumann levels. It is seen that position elements (which are the same for all excited states, as discussed before) at the Tsallis factor smaller than unity are strictly greater than their wave vector fellows crossing with them at $\alpha=1$ only. Compared to the ground level, the gap between position and momentum parts is wider and the latter ones are almost equal to each other for the quantum indices $n\gtrsim3$, as these were the cases for the Dirichlet well too. Thus, contrary to the R\'{e}nyi entropies, different values of the critical parameters $\alpha_{TH}$ do not alter qualitatively the shape of the Tsallis uncertainty relation.

\section{Conclusions}
A comparative analysis of the Dirichlet and Neumann quantum wells of width $a$ revealed similarities and differences between one-parameter measures $R_{\rho,\gamma}(\alpha)$ and $T_{\rho,\gamma}(\alpha)$ of these structures. Position functionals for both BCs are the same except the ground zero-energy Neumann level whose, e.g., R\'{e}nyi entropy is an $\alpha$-independent constant $\ln a$ what is explained by the flatness of the associated waveform. A crucial distinction between the two interface requirements is a difference between lower edges $\alpha_{TH}$ of the semi-infinite range where momentum entropies are defined, which are one quarter for the Dirichlet well and one half for the Neumann (and mixed Dirichlet-Neumann) one. Upon approaching this threshold, R\'{e}nyi measure logarithmically diverges. As a straight consequence of the gap between the two critical values, a sum of the position and momentum entropies entering uncertainty relation, Equation~\eqref{RenyiUncertainty2}, tends at large R\'{e}nyi coefficient to the finite Dirichlet level-dependent limit and unrestrictedly grows for the Neumann states. Both geometries support the earlier conjecture \cite{Olendski1} stating that the lowest-energy orbital of the magnetic field-free quantum system converts at $\alpha=1/2$ R\'{e}nyi and Tsallis uncertainty relations into the equality.

Dirichlet and Neumann BCs are limiting cases $\Lambda=0$ and $\Lambda=\infty$, respectively, of the Robin requirement \cite{Gustafson1}:
\begin{equation}\label{Robin1}
\left.{\bf n}{\bm\nabla}\Psi(\bf r)\right|_{\cal S}=\frac{1}{\Lambda}\Psi(\bf r)|_{\cal S},
\end{equation}
where $\bf n$ is an inward normal to the surface $\cal S$ and the length $\Lambda$ in general can take complex values \cite{Olendski5,Olendski6}. Recently, Shannon entropies $S_{\rho,\gamma}$, Fisher informations $I_{\rho,\gamma}$, Onicescu energies $O_{\rho,\gamma}$ and complexities $e^SO$ were scrutinized for the Robin well with real $\Lambda$ \cite{Olendski4}. Combining position and momentum functions derived in that research with the methods of analysis of the one-parameter measures developed above, one will be able to calculate R\'{e}nyi and Tsallis entropies of this structure. Preliminary, applying a convergence test to the momentum waveforms from Reference~\cite{Olendski4}, one sees that the critical threshold for any non-Dirichlet, $\Lambda\neq0$, well is just one half, as is, in particular, the case for the Neumann system, Equation~\eqref{NeumannThreshold1}, or attractive Robin wall \cite{Olendski1}. This singles out the $\Lambda=0$ BC from all other surface conditions. Detailed analysis of all properties of the R\'{e}nyi and Tsallis measures of the Robin well requires a separate careful investigation.

\section{Acknowledgements}
Research was supported by Competitive Research Project No. 2002143087 from the Research Funding Department, Vice Chancellor for Research and Graduate Studies, University of Sharjah.

\bibliographystyle{}

\end{document}